\RequirePackage{lineno}
\documentclass[twocolumn,letterpaper,aps,prd,superscriptaddress,amsmath,amssymb,showpacs]{revtex4}
\pdfoutput=1

\usepackage{xspace}
\usepackage{color}
\usepackage{graphicx}
\usepackage{dcolumn}
\usepackage{bm}
\usepackage{epstopdf}
\usepackage[
  bookmarks%
  ,pdftitle={Search for diboson production in met+bbbar at Tevatron}%
  ,pdfsubject={CDF Run 2 physics result}%
  ,pdfauthor={CDF Collaboration}%
  ,pdfstartview={Fit}%
  ]{hyperref}
\usepackage[all]{hypcap}
\usepackage{multirow}
\DeclareGraphicsExtensions{.pdf,.eps,.png,.jpg,.jpeg}
\graphicspath{{figures/},{figures/controlregions_20110323/},{figures/JES/},{figures/JES_fits/}}

\newcommand{\ppbar}{\ensuremath{p\overline{p}}\xspace}

\def\mMEt{\not\kern-.35em {E_T}}

\newcommand{\mett}{\mbox{$E\!\!\!\!/_{T}$}\xspace}
\newcommand{\mettvec}{\mbox{$\vec{E}\!\!\!\!/_{T}$}\xspace}
\newcommand{\met}{\mett}
\newcommand{\mpt}{\mbox{$P\!\!\!\!/_{T}$}}

\newcommand{\invfb}{\ensuremath{\rm fb^{-1}}\,}
\def\simle{\mathrel{
   \rlap{\raise 0.511ex \hbox{$<$}}{\lower 0.511ex \hbox{$\sim$}}}}

\begin{document}

\mbox{FERMILAB-PUB-11-375-E-PPD}
\hfill\mbox{CDF/PHYS/ELECTROWEAK/PUBLIC/10311}


\title{Search for $WZ+ZZ$ production with $\met+{}$jets with $b$
  enhancement at $\sqrt{s}=1.96$~TeV}

\affiliation{Institute of Physics, Academia Sinica, Taipei, Taiwan 11529, Republic of China} 
\affiliation{Argonne National Laboratory, Argonne, Illinois 60439, USA} 
\affiliation{University of Athens, 157 71 Athens, Greece} 
\affiliation{Institut de Fisica d'Altes Energies, ICREA, Universitat Autonoma de Barcelona, E-08193, Bellaterra (Barcelona), Spain} 
\affiliation{Baylor University, Waco, Texas 76798, USA} 
\affiliation{Istituto Nazionale di Fisica Nucleare Bologna, $^{aa}$University of Bologna, I-40127 Bologna, Italy} 
\affiliation{University of California, Davis, Davis, California 95616, USA} 
\affiliation{University of California, Los Angeles, Los Angeles, California 90024, USA} 
\affiliation{Instituto de Fisica de Cantabria, CSIC-University of Cantabria, 39005 Santander, Spain} 
\affiliation{Carnegie Mellon University, Pittsburgh, Pennsylvania 15213, USA} 
\affiliation{Enrico Fermi Institute, University of Chicago, Chicago, Illinois 60637, USA}
\affiliation{Comenius University, 842 48 Bratislava, Slovakia; Institute of Experimental Physics, 040 01 Kosice, Slovakia} 
\affiliation{Joint Institute for Nuclear Research, RU-141980 Dubna, Russia} 
\affiliation{Duke University, Durham, North Carolina 27708, USA} 
\affiliation{Fermi National Accelerator Laboratory, Batavia, Illinois 60510, USA} 
\affiliation{University of Florida, Gainesville, Florida 32611, USA} 
\affiliation{Laboratori Nazionali di Frascati, Istituto Nazionale di Fisica Nucleare, I-00044 Frascati, Italy} 
\affiliation{University of Geneva, CH-1211 Geneva 4, Switzerland} 
\affiliation{Glasgow University, Glasgow G12 8QQ, United Kingdom} 
\affiliation{Harvard University, Cambridge, Massachusetts 02138, USA} 
\affiliation{Division of High Energy Physics, Department of Physics, University of Helsinki and Helsinki Institute of Physics, FIN-00014, Helsinki, Finland} 
\affiliation{University of Illinois, Urbana, Illinois 61801, USA} 
\affiliation{The Johns Hopkins University, Baltimore, Maryland 21218, USA} 
\affiliation{Institut f\"{u}r Experimentelle Kernphysik, Karlsruhe Institute of Technology, D-76131 Karlsruhe, Germany} 
\affiliation{Center for High Energy Physics: Kyungpook National University, Daegu 702-701, Korea; Seoul National University, Seoul 151-742, Korea; Sungkyunkwan University, Suwon 440-746, Korea; Korea Institute of Science and Technology Information, Daejeon 305-806, Korea; Chonnam National University, Gwangju 500-757, Korea; Chonbuk National University, Jeonju 561-756, Korea} 
\affiliation{Ernest Orlando Lawrence Berkeley National Laboratory, Berkeley, California 94720, USA} 
\affiliation{University of Liverpool, Liverpool L69 7ZE, United Kingdom} 
\affiliation{University College London, London WC1E 6BT, United Kingdom} 
\affiliation{Centro de Investigaciones Energeticas Medioambientales y Tecnologicas, E-28040 Madrid, Spain} 
\affiliation{Massachusetts Institute of Technology, Cambridge, Massachusetts 02139, USA} 
\affiliation{Institute of Particle Physics: McGill University, Montr\'{e}al, Qu\'{e}bec, Canada H3A~2T8; Simon Fraser University, Burnaby, British Columbia, Canada V5A~1S6; University of Toronto, Toronto, Ontario, Canada M5S~1A7; and TRIUMF, Vancouver, British Columbia, Canada V6T~2A3} 
\affiliation{University of Michigan, Ann Arbor, Michigan 48109, USA} 
\affiliation{Michigan State University, East Lansing, Michigan 48824, USA}
\affiliation{Institution for Theoretical and Experimental Physics, ITEP, Moscow 117259, Russia}
\affiliation{University of New Mexico, Albuquerque, New Mexico 87131, USA} 
\affiliation{Northwestern University, Evanston, Illinois 60208, USA} 
\affiliation{The Ohio State University, Columbus, Ohio 43210, USA} 
\affiliation{Okayama University, Okayama 700-8530, Japan} 
\affiliation{Osaka City University, Osaka 588, Japan} 
\affiliation{University of Oxford, Oxford OX1 3RH, United Kingdom} 
\affiliation{Istituto Nazionale di Fisica Nucleare, Sezione di Padova-Trento, $^{bb}$University of Padova, I-35131 Padova, Italy} 
\affiliation{LPNHE, Universite Pierre et Marie Curie/IN2P3-CNRS, UMR7585, Paris, F-75252 France} 
\affiliation{University of Pennsylvania, Philadelphia, Pennsylvania 19104, USA}
\affiliation{Istituto Nazionale di Fisica Nucleare Pisa, $^{cc}$University of Pisa, $^{dd}$University of Siena and $^{ee}$Scuola Normale Superiore, I-56127 Pisa, Italy} 
\affiliation{University of Pittsburgh, Pittsburgh, Pennsylvania 15260, USA} 
\affiliation{Purdue University, West Lafayette, Indiana 47907, USA} 
\affiliation{University of Rochester, Rochester, New York 14627, USA} 
\affiliation{The Rockefeller University, New York, New York 10065, USA} 
\affiliation{Istituto Nazionale di Fisica Nucleare, Sezione di Roma 1, $^{ff}$Sapienza Universit\`{a} di Roma, I-00185 Roma, Italy} 

\affiliation{Rutgers University, Piscataway, New Jersey 08855, USA} 
\affiliation{Texas A\&M University, College Station, Texas 77843, USA} 
\affiliation{Istituto Nazionale di Fisica Nucleare Trieste/Udine, I-34100 Trieste, $^{gg}$University of Udine, I-33100 Udine, Italy} 
\affiliation{University of Tsukuba, Tsukuba, Ibaraki 305, Japan} 
\affiliation{Tufts University, Medford, Massachusetts 02155, USA} 
\affiliation{University of Virginia, Charlottesville, Virginia 22906, USA}
\affiliation{Waseda University, Tokyo 169, Japan} 
\affiliation{Wayne State University, Detroit, Michigan 48201, USA} 
\affiliation{University of Wisconsin, Madison, Wisconsin 53706, USA} 
\affiliation{Yale University, New Haven, Connecticut 06520, USA} 
\author{T.~Aaltonen}
\affiliation{Division of High Energy Physics, Department of Physics, University of Helsinki and Helsinki Institute of Physics, FIN-00014, Helsinki, Finland}
\author{B.~\'{A}lvarez~Gonz\'{a}lez$^w$}
\affiliation{Instituto de Fisica de Cantabria, CSIC-University of Cantabria, 39005 Santander, Spain}
\author{S.~Amerio}
\affiliation{Istituto Nazionale di Fisica Nucleare, Sezione di Padova-Trento, $^{bb}$University of Padova, I-35131 Padova, Italy} 

\author{D.~Amidei}
\affiliation{University of Michigan, Ann Arbor, Michigan 48109, USA}
\author{A.~Anastassov}
\affiliation{Northwestern University, Evanston, Illinois 60208, USA}
\author{A.~Annovi}
\affiliation{Laboratori Nazionali di Frascati, Istituto Nazionale di Fisica Nucleare, I-00044 Frascati, Italy}
\author{J.~Antos}
\affiliation{Comenius University, 842 48 Bratislava, Slovakia; Institute of Experimental Physics, 040 01 Kosice, Slovakia}
\author{G.~Apollinari}
\affiliation{Fermi National Accelerator Laboratory, Batavia, Illinois 60510, USA}
\author{J.A.~Appel}
\affiliation{Fermi National Accelerator Laboratory, Batavia, Illinois 60510, USA}
\author{A.~Apresyan}
\affiliation{Purdue University, West Lafayette, Indiana 47907, USA}
% ...
\collaboration{CDF Collaboration\footnote{With visitors from $^a$Istituto Nazionale di Fisica Nucleare, Sezione di Cagliari, 09042 Monserrato (Cagliari), Italy,
$^b$University of CA Irvine, Irvine, CA  92697, USA,
$^c$University of CA Santa Barbara, Santa Barbara, CA 93106, USA,
$^d$University of CA Santa Cruz, Santa Cruz, CA  95064, USA,
$^e$CERN,CH-1211 Geneva, Switzerland,
$^f$Cornell University, Ithaca, NY  14853, USA, 
$^g$University of Cyprus, Nicosia CY-1678, Cyprus, 
$^h$Office of Science, U.S. Department of Energy, Washington, DC 20585, USA,
$^i$University College Dublin, Dublin 4, Ireland,
$^j$University of Fukui, Fukui City, Fukui Prefecture, Japan 910-0017,
$^k$Universidad Iberoamericana, Mexico D.F., Mexico,
$^l$Iowa State University, Ames, IA  50011, USA,
$^m$University of Iowa, Iowa City, IA  52242, USA,
$^n$Kinki University, Higashi-Osaka City, Japan 577-8502,
$^o$Kansas State University, Manhattan, KS 66506, USA,
$^p$University of Manchester, Manchester M13 9PL, United Kingdom,
$^q$Queen Mary, University of London, London, E1 4NS, United Kingdom,
$^r$University of Melbourne, Victoria 3010, Australia,
$^s$Muons, Inc., Batavia, IL 60510, USA,
$^t$Nagasaki Institute of Applied Science, Nagasaki, Japan, 
$^u$National Research Nuclear University, Moscow, Russia,
$^v$University of Notre Dame, Notre Dame, IN 46556, USA,
$^w$Universidad de Oviedo, E-33007 Oviedo, Spain, 
$^x$Texas Tech University, Lubbock, TX  79609, USA,
$^y$Universidad Tecnica Federico Santa Maria, 110v Valparaiso, Chile,
$^z$Yarmouk University, Irbid 211-63, Jordan,
$^{hh}$On leave from J.~Stefan Institute, Ljubljana, Slovenia, 
}}
\noaffiliation

\date{October 13, 2011}

\begin{abstract}
  Diboson production ($WW+WZ+ZZ$) has been observed at the Tevatron in
  hadronic decay modes dominated by the $WW$ process.  This paper
  describes the measurement of the cross section of $WZ$ and $ZZ$
  events in final states with large $\mett$ and using $b$-jet
  identification as a tool to suppress $WW$ contributions. Due to the
  limited energy resolution, we cannot distinguish between partially
  hadronic decays of $WZ$ and $ZZ$, and we measure the sum of these
  processes. %
  The number of signal events is extracted using a simultaneous fit to
  the invariant mass distribution of the two jets for events with two
  $b$-jet candidates and events with fewer than two $b$-jet
  candidates.  We measure a cross section $\sigma(p\bar{p}\to WZ,ZZ) =
  5.8^{+3.6}_{-3.0}$~pb, in agreement with the standard model.
\end{abstract}
  
  
\maketitle
  


\section{Introduction}
\label{sec:Intro}
Measurements of diboson production cross sections provide tests of the
self-interactions of the gauge bosons. Deviations from the standard
model (SM) prediction for the production rates could indicate new
physics~\cite{hagiwara,kober}, specifically in hadronic final states~\cite{technicolor}. 
Furthermore, given that hadronic final states in diboson
production are similar to associated Higgs boson
production (Higgs-strahlung), $p\bar{p}\to VH+X$\,($V$=$W,Z$), the analysis
techniques described in this Letter are important for Higgs boson
searches~\cite{higgs_metbb}.

Diboson production has been observed at the Tevatron in fully leptonic
final states~\cite{cdf_ww_dil, d0_ww_dil}. In the case of partially
hadronic decay modes, the CDF collaboration observed a signal for
combined measurement of $WW$, $WZ$, and $ZZ$ using an integrated
luminosity of 3.5\,\invfb where the
signal is dominated by $WW$~\cite{metjj_prl, lmetjj_prl}. In this
paper, we describe a measurement where we isolate the $WZ$ and $ZZ$
signals in partially hadronic decay channels by requiring the presence
of $b$-jet candidates. We perform a fit to the dijet invariant mass
spectrum ($m_{jj}$), splitting events into two non-overlapping
classes: with at least two $b$-jet candidates (two-tag channel), and
fewer than two $b$-jet candidates (no-tag channel)~\cite{wwcomment}. This ensures
maximum acceptance to the $WZ+ZZ$ events, and fitting in both the
two-tag and the no-tag channel improves our signal sensitivity
significantly compared to using only one channel (with or without
$b$-tagging).  The signatures to which we are sensitive are $WZ\to \ell
\nu b\bar{b}$ and $ZZ \to \nu \bar{\nu} b \bar{b}$ in the two-tag channel
and all decays with unbalanced transverse momentum ($\mett$) in the
no-tag channel ($WZ\to \ell \nu q\bar{q}, q\bar{q}^\prime\nu\bar{\nu}$ and
$ZZ \to \nu \bar{\nu} q \bar{q}$)~\cite{metdef}.


\section{The CDF Detector}
The CDF~II detector is described in detail
elsewhere~\cite{CDF_detect_A}.  The detector is cylindrically
symmetric around the proton beam axis which is oriented in the
positive $z$ direction.  The polar angle, $\theta$, is measured from
the origin of the coordinate system at the center of the detector with
respect to the $z$ axis. Pseudorapidity, transverse energy, and
transverse momentum are defined as $\eta$=$-\ln\tan(\theta/2)$,
$E_{T}$=$E\sin\theta$, and $p_{T}$=$p\sin\theta$, respectively.  The
central and plug calorimeters, which respectively cover the
pseudorapidity regions of $|\mathrm\eta|$$<$1.1 and
1.1$<$$|\mathrm\eta|$$<$3.6, surround the tracking system with a
projective tower geometry. The detector has a charged particle
tracking system immersed in a 1.4~T magnetic field, aligned coaxially
with the $\ppbar$ beams.  A silicon microstrip detector provides
tracking over the radial range 1.5 to 28~cm.  A 3.1~m long open-cell
drift chamber, the central outer tracker (COT), covers the radial
range from 40 to 137~cm and provides up to 96 measurements with
alternating axial and $\pm2^{\circ}$ stereo superlayers.  The fiducial
region of the silicon detector extends to $|\eta|\sim 2$, while the
COT provides coverage for $|\eta|\simle 1$.
Muons are detected up to
$|\eta| < 1.0$ by drift chambers located outside the hadronic
calorimeters.

\section{Dataset and Event Selection}
We analyze a dataset of $\ppbar$ collisions corresponding to an
integrated luminosity of 5.2\,\invfb collected with the CDF~II
detector at a center-of-mass energy of 1.96\,TeV.
Events are selected via a set of triggers with \mett requirements.
The bulk of the data is collected with a trigger
threshold $\mett>45$~GeV. Other triggers have a lower \mett
requirement but also include additional requirements on jets in the
event, or sometimes correspond to smaller effective integrated
luminosity.  
We measure the trigger efficiency using an independent $Z\to\mu\mu$
sample and verify that the trigger logic used does not sculpt the shape of the dijet invariant mass. 


Events with large \mett ($\mett>50~\mathrm{GeV}$) and two or more jets are selected in this analysis.
Jets are reconstructed in the calorimeter using the {\sc jetclu} cone
algorithm~\cite{jetclu} with a cone radius of 0.4 in ($\eta,\phi$)
space. The energy measured by the calorimeter is corrected for effects
that distort the true jet 
energy~\cite{jesnim}. Such effects include the non-linear response of the calorimeter to
particle energy, loss of energy in uninstrumented regions of the
detector, energy radiated outside of the jet cone, and multiple
proton antiproton interactions per beam crossing. 
The jets must have $E_T>20$~GeV and be within $|\eta|<2$.
To suppress the
multi-jet background contribution, we require the azimuthal angle between the $\mett$
vector and any identified jet, $\Delta\phi(\mett,\mathrm{jet})$, to be
larger than 0.4 radians~\cite{whichjets}. 
The $\mett$-significance, as defined in \cite{metjj_prl}, measures the likelihood that the $\mett$ in the
event comes from actual particles escaping detection as opposed to
resolution effects and is typically low when $\mett$ arises from
mis-measurements. We require $\mett$-significance to be larger than 4
(see \cite{metjj_prl, emtiming}). Beam halo events are removed by requiring
the event electromagnetic fraction, defined as the ratio between the
amount of energy measured in the electromagnetic calorimeter and the
sum of electromagnetic and hadronic calorimeter measurements,
$E_\mathrm{EM}/E_\mathrm{total}$, to be between 0.3 and 0.85.  We
remove cosmic ray events based on timing information from the
electromagnetic and hadronic calorimeters.


\section{Selecting $b$ quark jets}
To gain sensitivity to the $b$-quark content of our jet sample, we
employ a new multivariate neural network based tagger that provides a
figure of merit to indicate how $b$-like a jet appears to be.
This tagger is unique in its emphasis on studying individual tracks.
A more detailed description of this tagger may be found in \cite{nim_bness}.
The tagger identifies tracks with transverse momentum $p_{T} > 0.4~\mathrm{GeV/c}$
that have registered hits in the innermost (silicon) tracking layers,
and uses a track-by-track neural network to calculate a figure of
merit for a given track's ``$b$ness'', \emph{i.e.}, the likelihood
that it comes from the decay of a $B$ hadron. 
The observables used in the track neural network are the
transverse momentum of the track in the laboratory frame, the transverse
momentum of the track with respect to the jet axis, the rapidity with
respect to the jet axis and the track impact parameter with respect to
the primary vertex and its uncertainty. The output of the track neural
network is a numerical value in the range from -1 to 1.

Having the track $b$nesses, we proceed to calculate the jet-by-jet
$b$nesses. We use tracks with track-by-track NN values greater than
-0.5 in the fitting of a secondary vertex. The observables used as
inputs to the jet neural network are the top five track $b$nesses in
the jet cone, the number of tracks with positive track $b$ness, the
significance~\cite{signif} of the displacement of the secondary vertex
from the $B$-hadron decay in the $xy$ plane, the invariant mass of the
tracks used to fit the displaced vertex, the number of $K_S$
candidates found in the jet, and muon information for semileptonic $B$
decays as described in ~\cite{cdf_smt}.  We include the number of
$K_S$ candidates found since a much higher fraction of $b$ jets than
non-$b$ jets contain $K_S$ particles.  The final output of the
algorithm is a number between -1 and 1, the $b$ness. By requiring
values of $b$ness closer to 1, one can select increasingly pure
samples of $b$ jets.  The training for the track neural network as
well as the jet-by-jet network is performed using jets matched to $b$
quarks from $Z\rightarrow b\bar{b}$ events for signal and jets not
matched to $b$ quarks for background in a {\sc pythia} $ZZ$ Monte
Carlo sample.

To verify that the $b$-tagger data response is reproduced by the Monte
Carlo simulation, we use two control samples, one dominated by
$Z(\to\ell\ell)$ + 1 jet events, and one dominated by $t\bar{t}$
pair events using a lepton + jets selection.  The former offers a
comparison of jets that largely do not originate from bottom quarks,
while the latter compares jets in a heavily $b$-enhanced
sample.  
We examine the $b$ness distributions in simulation and data and use
these comparisons to derive a correction to the tagging efficiency and
mistag rates, the rate of misidentification of
non-$b$ jets as $b$ jets, in the Monte Carlo simulation for the cuts 
on the jet $b$ness that define our tagged selection. 
The operating point of our $b$ tagger utilizes a tight cut on the highest $b$ness 
jet in the event, and a looser cut on the second highest $b$ness jet. We list the tagging 
efficiencies and mistag rates for these cuts in Table~\ref{tab:btag-values}. 
Further details of their determination are in~\cite{nim_bness}.
We correct the MC, as it underestimates the observed mistag rate and 
overestimates the observed efficiency.

\begin{table}[hbt]
\begin{center}
\begin{tabular}{ll|cc} \hline
 & & Data & Scale Factor on MC  \\ \hline 
Mistag Rate & 1\textsuperscript{st} jet & $1.00\pm0.21\%$ & $1.15\pm0.24$  \\ 
& 2\textsuperscript{nd} jet & $8.19\pm0.34\%$ & $1.14\pm0.05$ \\ \hline 
Tag Efficiency &  1\textsuperscript{st} jet & $65.2\pm4.0\%$ & $0.95\pm0.06$  \\ 
& 2\textsuperscript{nd} jet & $62.2\pm5.4\%$ & $0.91\pm0.08$  \\ \hline
\end{tabular}
\caption{Mistag rates and efficiencies on jet $b$ness cuts determined from comparisons of data and MC in our $Z +$ jet and $t\bar{t}$ control samples. As we order jets in $b$ness, the 1\textsuperscript{st} jet is the highest $b$ness jet in the event, and the 2\textsuperscript{nd} jet is the 2\textsuperscript{nd} highest $b$ness jet in the event. The MC tends to overestimate the tagging efficiency and underestimate the mistag rate, and so we apply a correction.}
\label{tab:btag-values}
\end{center}
\end{table}

\section{Background estimation}
We define our signal sample as events in the $40<m_{jj}<160~\mathrm{GeV}/c^2$
region.  
In the calculation of the invariant mass $m_{jj}$ we use the two jets
in the events with the highest $b$ness score.  The final number of
events is extracted by a simultaneous fit to the dijet invariant mass
distribution in the two-tag and no-tag channels, as defined
above. Since we apply $b$-tagging and allow for two or more jets,
$t\bar t$ and single $t$ production are a significant background.  To
further suppress these backgrounds, we require the events to have no
more than one identified lepton (electrons or muons), where a very
loose lepton identification is used to increase the efficiency of this
rejection. In addition, the sum of the number of identified electrons,
muons and jets with $E_T>10$~GeV must not exceed 4.

After this selection, we have four major classes of backgrounds:
\begin{enumerate}
\item Electroweak (EWK) $V$ boson+jet processes that are estimated using
  Monte Carlo simulations and cross-checked using a
  $\gamma$+jets data set, described below.
\item Multi-jet events with
  generic QCD jet production which result in \mett due to
  mis-measurements of the jet momenta. This background is evaluated
  using a data-driven method.
\item Single top and top quark pair production.
  We estimate this background using a Monte Carlo simulation.
\item $WW\rightarrow l\nu jj$ production. This is indistinguishable
  from the signal in the non-$b$-tagged region. This background is
  evaluated using a Monte Carlo simulation.
\end{enumerate}


Monte Carlo simulations used for signal and background estimates are performed
with a combination of {\sc pythia}~\cite{pythia}, {\sc
  alpgen}~\cite{alpgen} and {\sc MadGraph}~\cite{madgraph} event generators interfaced with {\sc
  pythia} 
for parton showering. The geometric and kinematic acceptances are
obtained using a {\sc geant}-based simulation of the CDF II
detector~\cite{geant3}. For the comparison to data, all sample cross sections are normalized to the
results of NLO  
calculations performed with {\sc mcfm~v5.4} program~\cite{campbell} and using the \textsc{cteq6m} parton
distribution functions  
(PDFs)~\cite{cteq6m}.

\subsection{Multi-jet background}
\label{sec:qcd}

Multi-jet production does not typically contain large intrinsic
$\met$.  The underlying assumption of how multi-jet background enters
the analysis is that either jets are mis-measured, or that a charged
or neutral hadron or a $\gamma$ is lost in an uninstrumented
region of the detector. We expect the dominant effect to be jet
mis-measurement. Because of the high cross section of multi-jet
production, this can be a significant background in a $\met$+jets
based analysis.  
We derive both the
normalization and the dijet mass shape of the multi-jet background from data.
The final measure of
the amount of multi-jet background will be determined from the extraction fit.

\begin{figure*}[tbhp]
  \centering
  \includegraphics[width=0.49\textwidth]{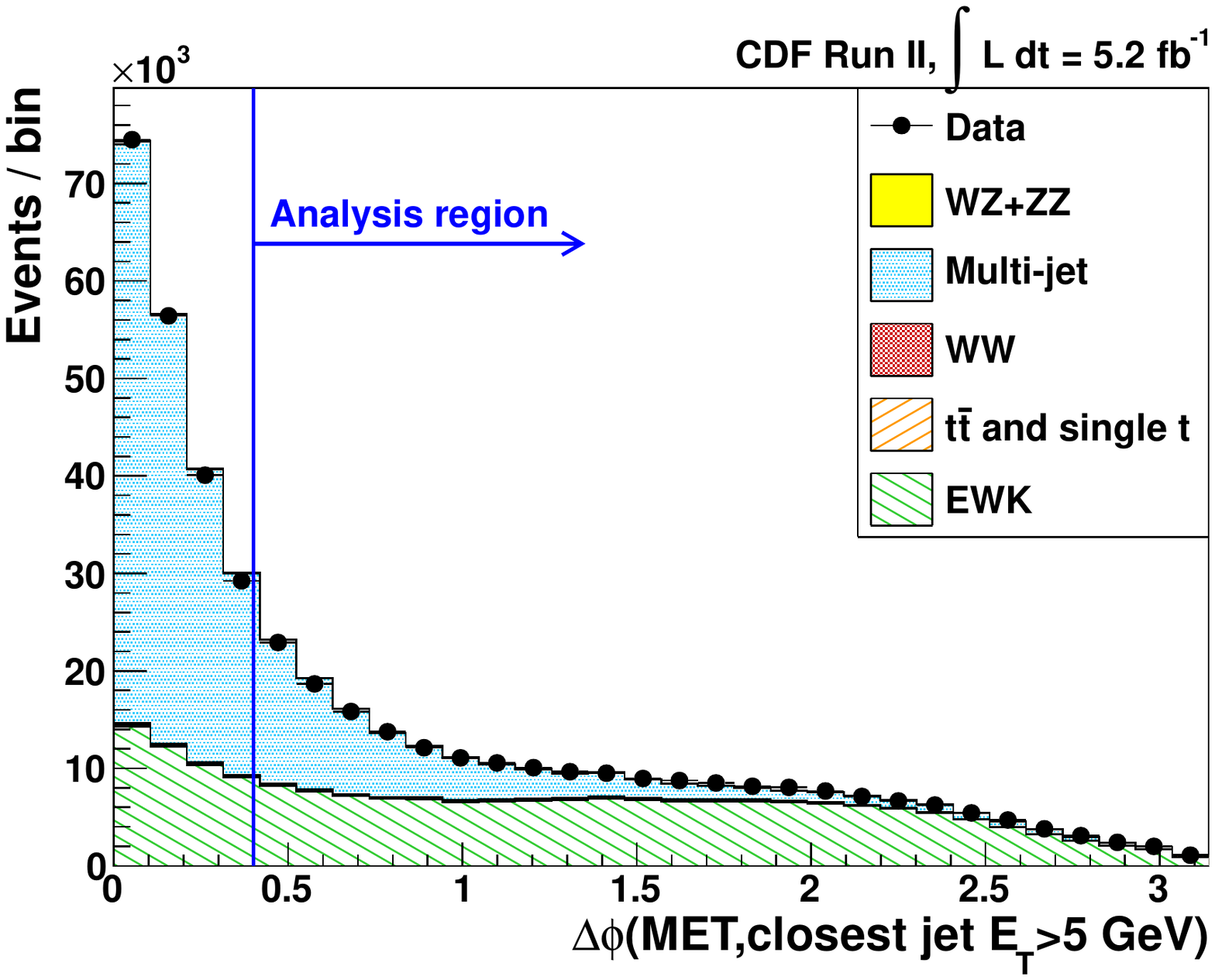}
  \includegraphics[width=0.49\textwidth]{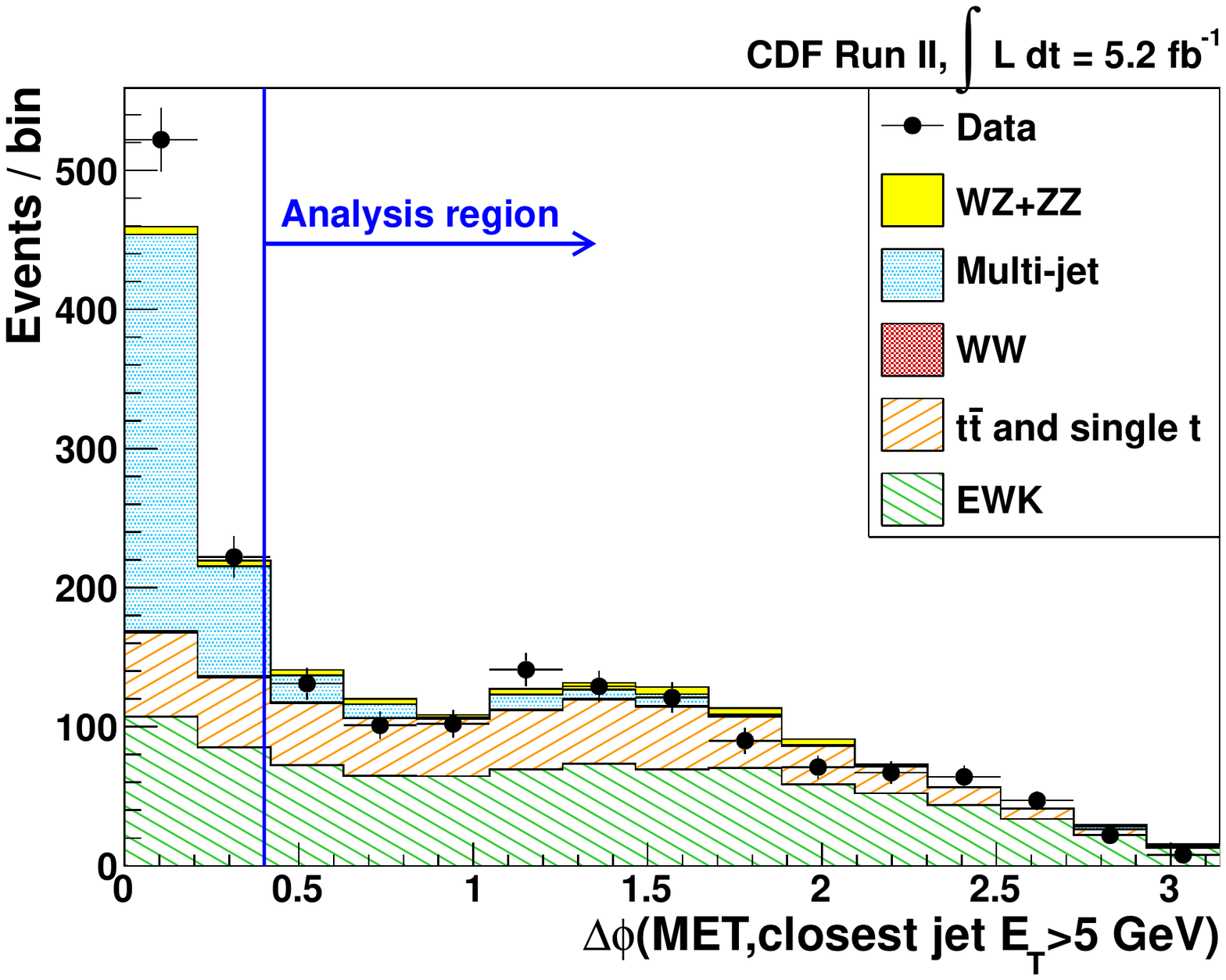}
  \includegraphics[width=0.49\textwidth]{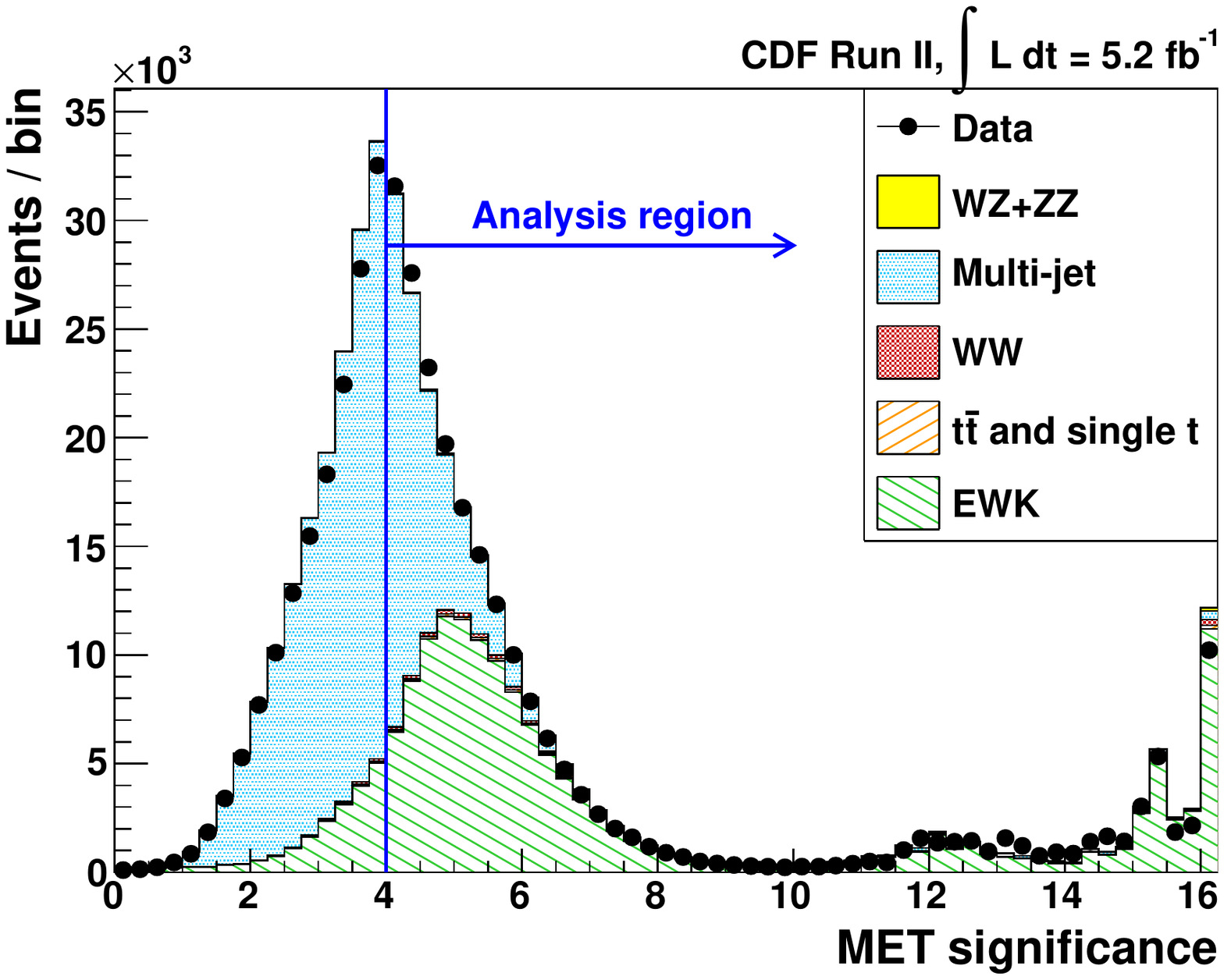}
  \includegraphics[width=0.49\textwidth]{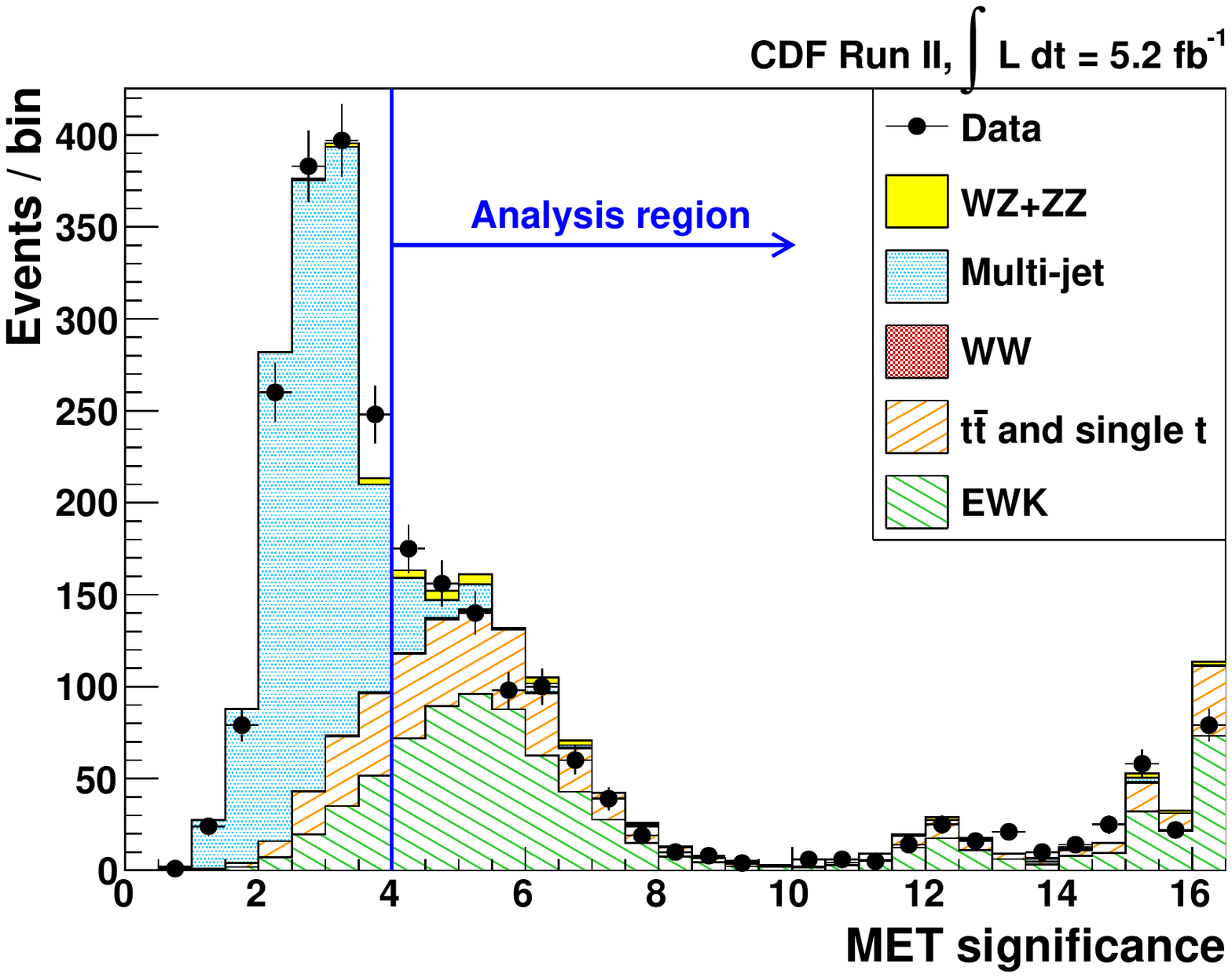}
  \caption{%
  {\it Left}: no-tag region. {\it Right}: 2-tag region.
  {\it Top row}: Minimum azimuthal angular separation
    min($\Delta\phi(\mettvec,\mathrm{jet})$) between all jets with
    $E_T>5$~GeV and the missing $E_T$, for events that pass all of the
    analysis cuts except for the min($\Delta\phi(\mettvec,\mathrm{jet})$)
    cut.  The analysis cut is at
    min($\Delta\phi(\mettvec,\mathrm{jet}))>0.4$.
  {\it Bottom row}: \mett-significance distribution for events
that pass all of the analysis cuts except for the \mett-significance
cut. The analysis cut is at \mett-significance${}>4$.
The highest bin is the overflow bin.
}
\label{fig:qcdtest}
\end{figure*}
The two important cuts used to reject this background are on the
$\mett$-significance and min($\Delta\phi(\mettvec,\mathrm{jet})$).
These distributions are shown in Fig.~\ref{fig:qcdtest}, which also
demonstrates our ability to model the multi-jet background.

To estimate the remaining multi-jet background contribution, we construct a new
variable, $\mpt$, to complement the traditional calorimeter-based
$\met$. The $\mpt$ is defined as the negative vector sum of tracks
with $p_T>0.3$~GeV/$c$.  Tracks used in the calculation of $\mpt$ have
to pass minimal quality requirements 
and be within a $\pm4\sigma$ window in the direction along the
beamline from the primary vertex.
  
\begin{figure}[tbp]
  \centering
  \includegraphics[width=0.5\textwidth]{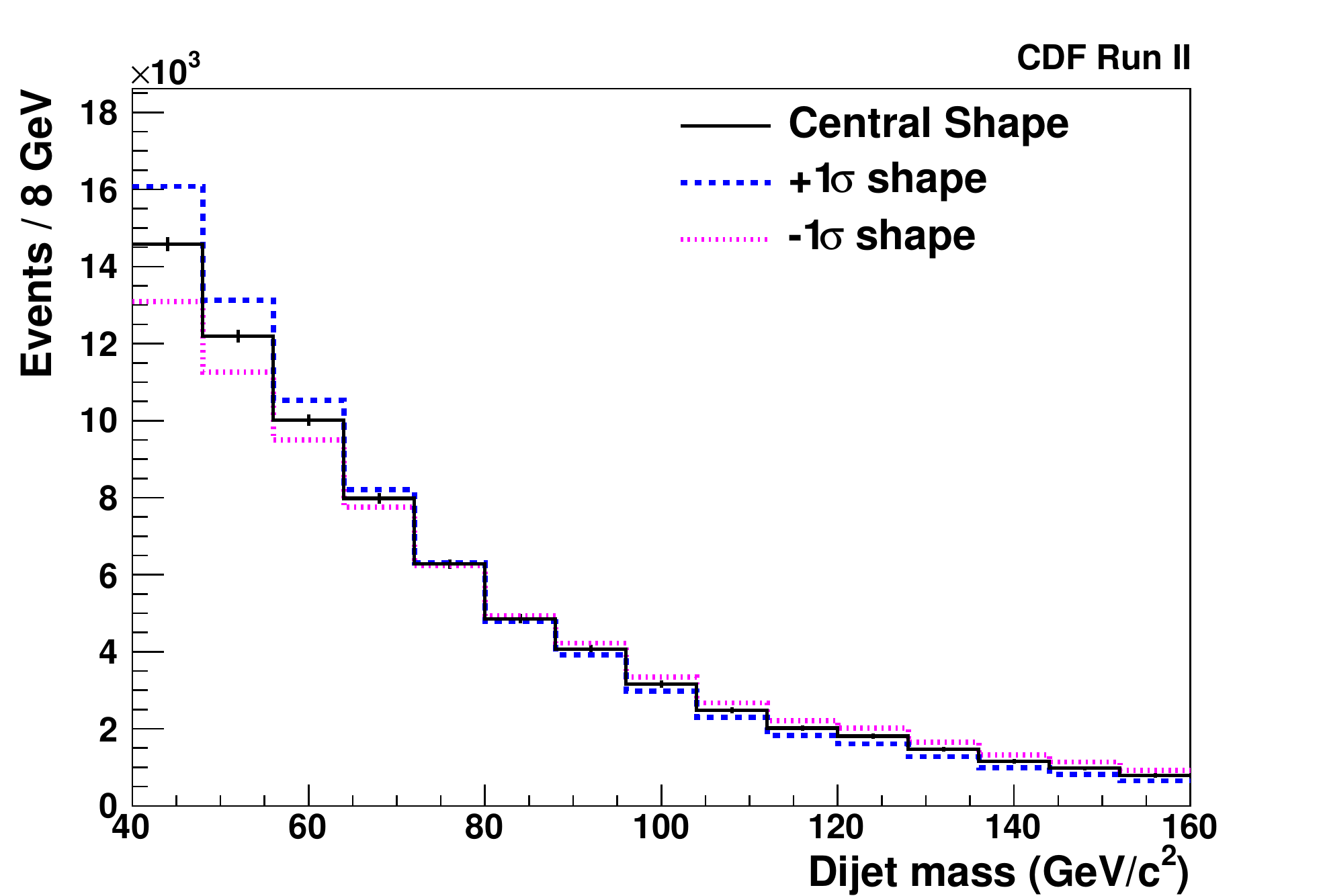}
  \caption{
    The multi-jet background dijet mass template and its
    corresponding shape uncertainties.
  }
\label{fig:qcdshape}
\end{figure}
When comparing the azimuthal angle ($\phi$) between $\met$ and $\mpt$,
we expect the two quantities to align in the case of true $\mett$
(\textit{e.g.}, for diboson signal and electroweak backgrounds).  The
difference between these two angles is referred to as
$\Delta\phi_{MET}$.
Electroweak backgrounds (and diboson 
signal) will be present in all regions, but will dominate at low 
$\Delta\phi_{MET}$ due to correctly measured $\met$ from neutrinos. 
To determine the dijet mass shape of the multi-jet background, we
subtract all other background predictions obtained with Monte Carlo
simulations from data, in the multi-jet enhanced region of
$\Delta\phi_{MET}>1$. The normalization of the template obtained this
way is then corrected to account for those events with
$\Delta\phi_{MET}\leq1$.  This correction introduces a 7\% uncertainty
on the normalization of the multi-jet background, where the
uncertainty was assessed by obtaining the correction factor both in
data and in a multi-jet Monte Carlo sample.  The uncertainty on the
shape of the distribution is estimated by comparing the difference in
dijet mass shapes for $\Delta\phi_{MET}>1$ and $\Delta\phi_{MET}<1$ in
a control sample defined by $3 < \mett$-significance $< 4$.  The
resulting multi-jet background dijet mass shape and its uncertainties
are shown in Fig.~\ref{fig:qcdshape} and are used as a shape
uncertainty in the fit.  For the two-tag channel we do not have enough
statistics to measure a shape, so we use the same shape as in the
no-tag region.


\subsection{Electroweak Shape Systematic}
\label{sec:gammajets}

Following the method used in the $\met+$jets analysis of~\cite{metjj_prl}, 
we use a $\gamma$+jets data sample to check our modeling of the
$V+$jet background shape. 
This is motivated by the similarities between the two types of processes.
While there are 
some differences (the $W$ and $Z$ bosons are massive, the photon is not, and 
unlike the $W$ the photon lacks charge), these are accounted for by a weighting 
procedure described below.

Along with differences in the physics, there are also differences in 
the detector response to $\gamma+$jets and $V+$jets. 
In order to have the  $\gamma+$jets events emulate the $V+$jets events, the photon $E_T$ is
vectorially subtracted from $\met$. Doing this, the $\gamma+$jets becomes topologically very
similar to the $Z+$jets with a $Z$ decaying to neutrinos, or a $W+$jets with a $W$ decaying to a
neutrino and a missed or poorly reconstructed lepton. 
A few other differences exist in the selection
cuts applied to $\gamma+$jets versus $\met+$jets data, shown in
Table~\ref{tab:photon_cuts}. As with the different approach to $\met$, these 
cuts are designed to allow for a data sample dominated by $\gamma+$jets
events and having adequate statistics.

\begin{table}[hbt]
\begin{center}
\begin{tabular}{|c|c|} \hline
$\met +$jets & $\gamma +$jets\\ \hline\hline
$\met > 50$\,GeV  &  $\bigl| \vec{\met} + \vec{E_T}_\text{photon}\bigr| > 50$\,GeV \\ 
$\Delta\phi(\vec{\met},\text{jet}) > 0.4$ & $\Delta\phi(\vec{\met} + \vec{E_T}_\text{photon},\text{jet}) > 0.4$ \\
$0.3 < \frac{\text{EM}}{E_\text{total}} < 0.85$ & $0.3 < \frac{\text{EM}}{E_\text{total}}$ \\ 
$\met$-\text{significance} $ > 4$ & -- \\
jet $b$ness cuts & -- \\
-- & $\gamma$ passes standard CDF cuts \\
-- & $\Delta R(\text{photon},\text{jet}) > 0.7$ \\
\hline


\end{tabular}
\caption{List of differences between cuts applied to the $\met + $ jets vs.
  $\gamma + $jets sample. A ``--'' denotes a lack of cut.}
\label{tab:photon_cuts}
\end{center}
\end{table}




\begin{figure*}[tbp]
  \centering
  \includegraphics[width=\textwidth]{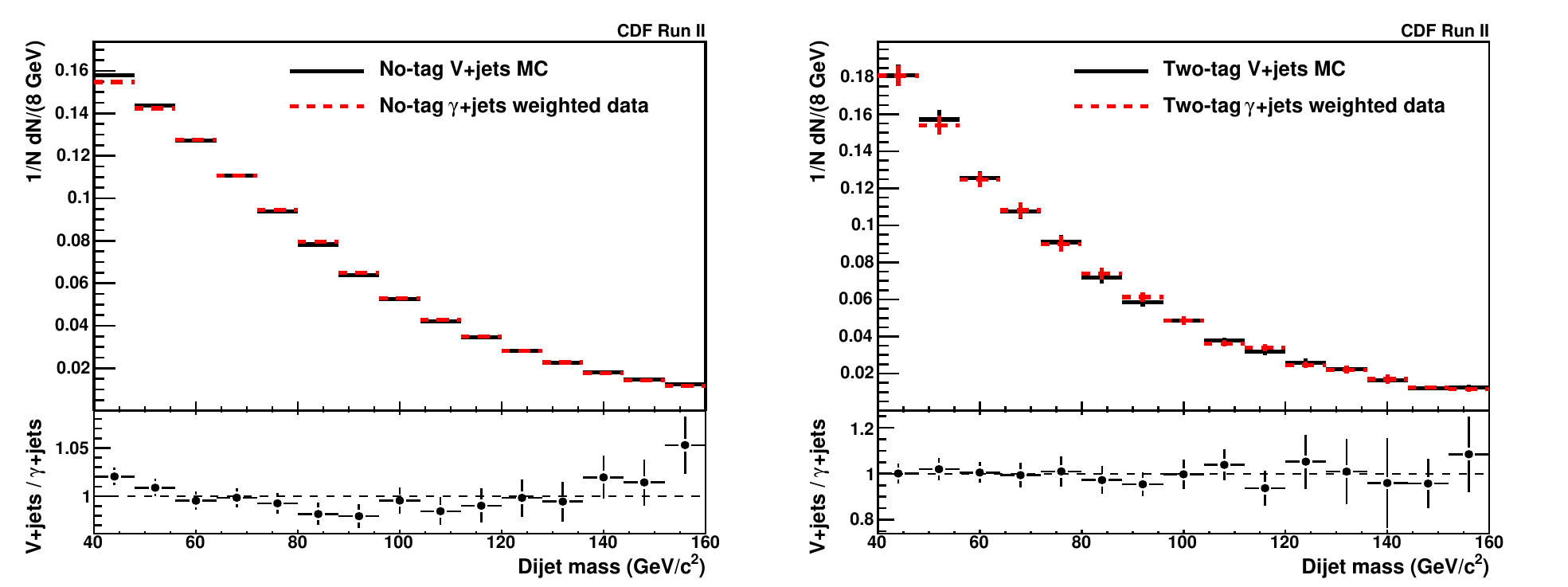}
\caption{Comparison of the $\gamma+$jets template with the electroweak
  MC template in the no-tag (left) and two-tag (right) regions.}
\label{fig:comp_ewk_vs_pho}
\end{figure*}
In order
to account for those remaining kinematic differences between $\gamma+$jets and
$V+$jets, we correct the $\gamma+$jets dijet mass shape in data based on the difference
between $\gamma+$jets and $V+$jets Monte Carlo simulations. 
First, 
the ratio of the $m_{jj}$ distributions from $V$+jets Monte Carlo simulation and 
inclusive $\gamma$+jets Monte Carlo simulation is obtained.
This ratio describes the difference 
in the physics of $\gamma+$jets and $V+$jets events. 
Note that since the $\gamma+$jets data sample will be contaminated with
$\gamma+W/Z \rightarrow\text{jets}$ events peaking in the signal region,
their expected contribution is subtracted from the $\gamma+$jets distribution.
Next, the $V$+jets / $\gamma+$jets $m_{jj}$ ratio histogram is 
multiplied bin-by-bin with the $\gamma+$jets data histogram, in effect sculpting the 
$\gamma+$jets data to look like $V+$jets data.
Since the Monte Carlo simulated events enter only in the ratio,
any production difference is taken into account while effects such as
detector resolution, PDF uncertainties and modeling of initial- and
final-state radiation cancel. 
After we apply this correction to the
$\gamma+$jets data, there is a residual difference, shown in
Fig.~\ref{fig:comp_ewk_vs_pho}, between the corrected
$\gamma+$jets data and our $V$+jets simulation, and we take this difference as a
systematic uncertainty on the shape of the $V$+jets background prediction.

\section{Signal extraction and results}
\label{sec:results}

\begin{figure*}[tbp]
	\centering
	\includegraphics[width=\textwidth]{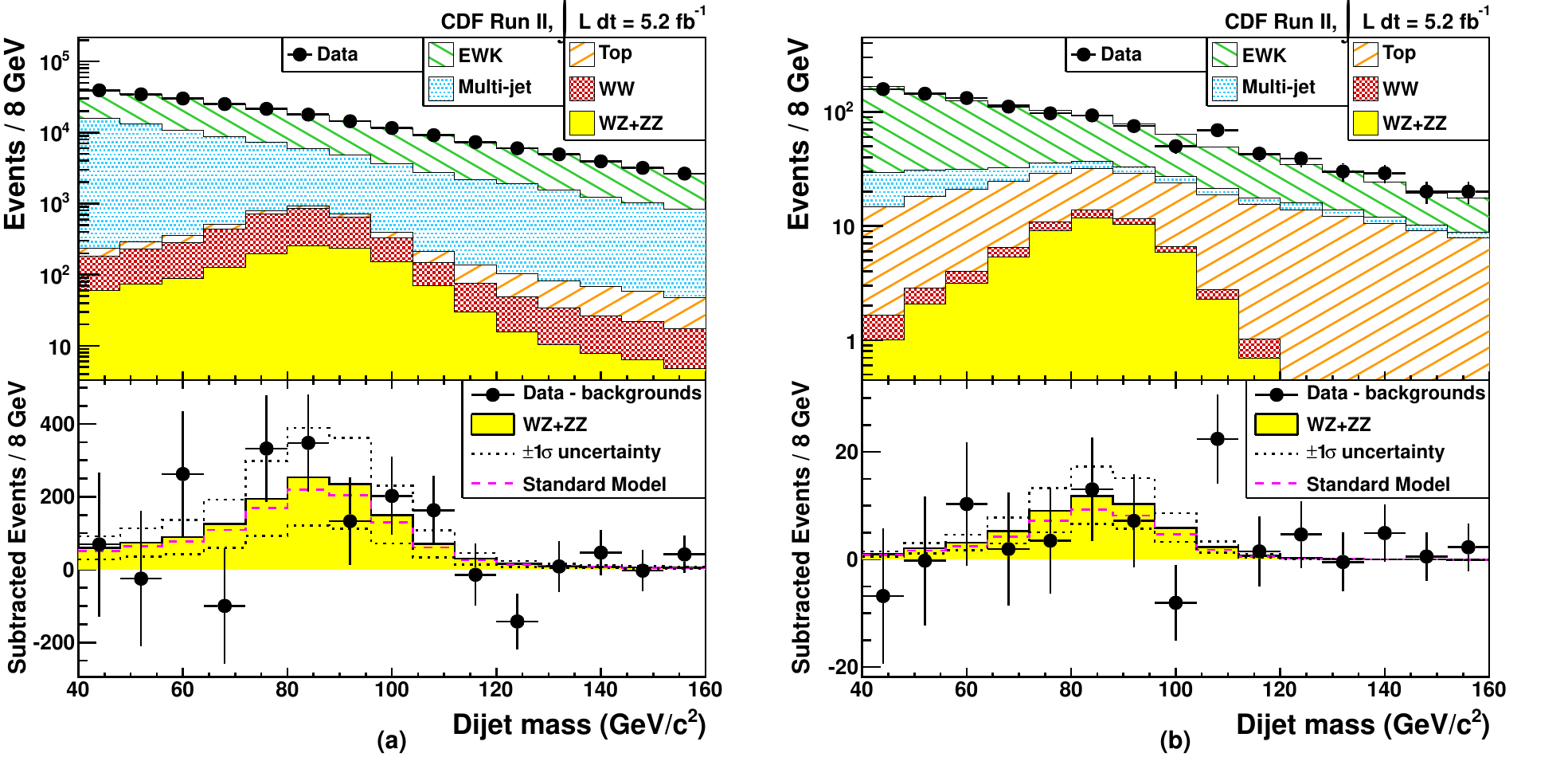}
        \caption{Result of the fit to data for the double fit to all
          of $WZ/ZZ$.  Left column is the no-tag channel; right column
          is the two-tag channel.  Bottom row shows data after the  background subtraction.
          }
\label{fig:results1}
\end{figure*}

We extract the number of signal events with a binned maximum likelihood fit to data using the
method described in \cite{mclimit}.  
We supply template histograms for
backgrounds and signals and perform a simultaneous fit in two  channels, defined by different $b$ness
thresholds.
The templates, and the uncertainties
on their normalizations, are listed below:
\begin{enumerate}
\item EWK background ($W/Z$+jets): Normalizations are allowed
  to float in the fit, unconstrained, with no correlation
  between the two tagging channels.
\item $t\bar t$ and single top: The uncertainties on the theoretical
  cross sections of these processes are 6\%~\cite{ttbar-xsec} and
  11\%~\cite{singlet-t,singlet-s}, respectively. We combine these two
  processes to a single template and treat these uncertainties as
  uncorrelated, which translates to an uncertainty of 5.8\%
  on the normalization of the no-tag channel template, and 5.4\% on
  the normalization of the two-tag channel template, due to the
  relative contributions of each process.
\item Multi-jet background: We use our data-driven estimate, Gaussian
  constrained with an uncertainty of 7\% in the no-tag
  channel. Because there are very few events in the two-tag channel
  template, we assign a normalization uncertainty equal to the
  statistical uncertainty ($\sqrt{N}/N$, 11\%) of the template. The
  uncertainties in the two channels are treated as uncorrelated.
\item $WW$: We use the NLO cross section and apply a Gaussian
  constraint to the number of $WW$ events centered on this value with
  a width equal to the theoretical uncertainty of 6\%~\cite{campbell}.
\item $WZ/ZZ$ signal: As this is our signal, its normalization is
  allowed to float unconstrained in the fit.  We assume that each
  signal process contributes proportionally to its predicted SM cross
  section: 3.6~pb for $WZ$ and 1.5~pb for $ZZ$ (\cite{campbell})
  corrected for our selection's acceptance and efficiencies.
\end{enumerate}

In addition to uncertainties on the normalizations of each template,
we consider other systematic uncertainties that may affect the shape
of templates. Shape uncertainties have been described for the
electroweak and multi-jet backgrounds previously. For top and diboson
samples, we consider the impact of the jet energy scale
and the effect that uncertainties due to the differences
between jet $b$ness behavior in data and Monte Carlo simulation may
have on the templates' shapes and normalizations.
These uncertainties are summarized in Table~\ref{tab:systematics}.
All of the above uncertainties are treated as
nuisance parameters and are incorporated into the fit using a Bayesian
marginalization technique~\cite{mclimit}.

\begin{table*}[tbp]
  \begin{center}
    \begin{tabular}{|cc|ccccc|} \hline
      \bf{Systematic Uncertainties} & channel & $WZ/ZZ$ & $WW$ & $t\bar{t}$ \& single $t$ & EWK & Multi-jet \\ \hline
      \multirow{2}{*}{Cross Section (Norm.)} & no-tag & \textit{Unconstr.} & $\pm$6\% & $\pm$5.8\% & \textit{Unconstr.} & $\pm$7\% \\
      & 2-tag & \textit{Unconstr.} & $\pm$6\% & $\pm$5.4\% & \textit{Unconstr.} & $\pm$11\% \\\hline
      \multirow{2}{*}{Jet Energy Scale} & no-tag & $\pm$7.1\% & $\pm$7.6\% &$\pm$2.2\%& & \\
      & 2-tag & $\pm$6.9\% & $\pm$7.6\% & $\pm$1.7\%& & \\\hline
      \multirow{2}{*}{$b$ness cuts (up)} & no-tag & $+0.46\%$ & $+0.08\%$ & $+3.0\%$ & & \\
      & 2-tag & $-13.0\%$ & $-24.2\%$ & $-11.8\%$ & & \\\hline
      \multirow{2}{*}{$b$ness cuts (down)} & no-tag & $-0.51\%$ & $-0.08\%$ & $-3.6\%$ & & \\
      & 2-tag & $+14.5\%$ & $+25.9\%$ & $+13.8\%$ & & \\ \hline 
    \end{tabular}
    \caption{A summary of the systematic uncertainties incorporated into the fit of the dijet mass distribution. The cross section normalizations of the signal and EWK templates are allowed to float in the fit, unconstrained. There are additional uncertainties on the shape of the EWK and Multi-jet templates, as described in the text. There is also an uncertainty on the shape of the diboson processes due to the jet energy scale. This shape uncertainty is correlated with the rate uncertainty shown here.
      }
\label{tab:systematics}
\end{center}
\end{table*}

We choose the jet $b$ness thresholds that define our two fitting
channels to optimize the significance of our final result.  The
optimization for the two-tag channel points to a broad region where
the sensitivity is maximized and we choose the operating point for our
$b$ness thresholds in that region. The optimization favors that all
remaining events be combined in a no-tag channel, rather than a
single-tag channel with a low $b$ness threshold.
\begin{table}[tbh]
  \begin{center}
    \begin{tabular}{|c|r@{$\,$}l|r@{$\,$}l|} \hline
      \multirow{2}{*}{Process} & \multicolumn{2}{c|}{Fit $N_\text{events}$}  & \multicolumn{2}{c|}{Fit $N_\text{events}$} \\
      & \multicolumn{2}{c|}{(no-tag)} &  \multicolumn{2}{c|}{(two-tag)} \\ \hline\hline
      EWK        & 149900 &$^{+5600}_{-5200}$ & 749 & $\pm48$\\
      $t\bar t$ and single $t$  & 898 & $^{+59}_{-61}$ & 217 &$^{+23}_{-27}$ \\
      Multi-jet  & 76600 &$^{+4900}_{-5300}$ & 76.3 & $\pm$9.0 \\
      $WW$         & 2720 & $\pm 200$ & 10.5  &$^{+2.1}_{-2.3}$ \\
      $WZ/ZZ$      & 1330 &$^{+710}_{-690}$ & 52 &$^{+24}_{-23}$ \\\hline
    \end{tabular}
    \caption{Extracted number of events from the 2-channel fit for $WZ$/$ZZ$,
     with all systematic uncertainties applied. Each uncertainty is reported to two significant figures, 
     and all event totals are reported to the precision reflected in the uncertainty.}
    \label{tab:nevt1}
  \end{center}
\end{table}
Figure~\ref{fig:results1} shows the results of the fit, and
Table~\ref{tab:nevt1} shows the number of fitted events.  

\begin{figure*}[tbp]
 	\centering
 	\includegraphics[width=0.8\textwidth]{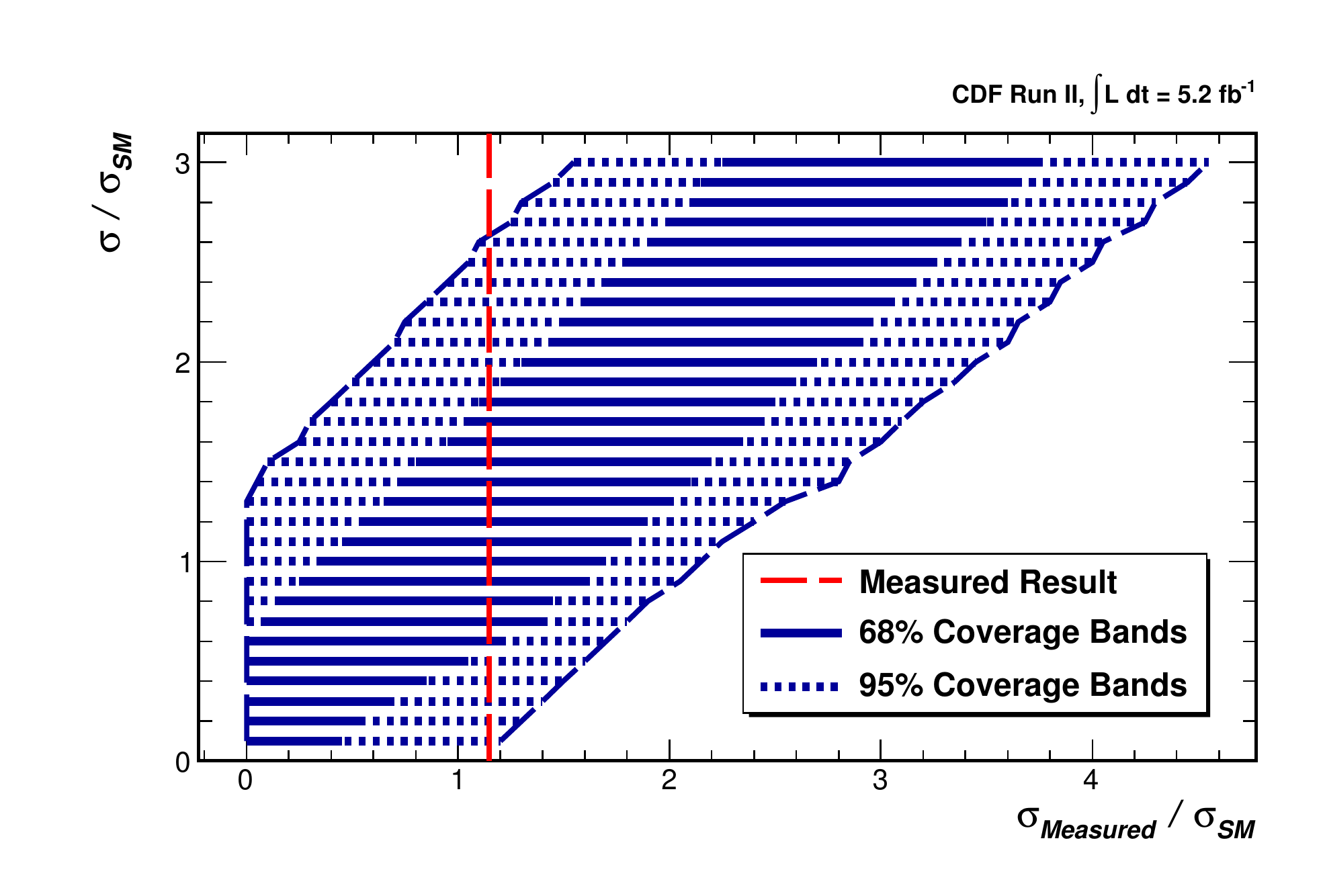}
         \caption{Confidence bands showing
           the expected range of measured cross sections as a function
           of the true cross section, with 68\% CL (blue solid region) and
           95\% CL (blue dotted region). Our measured result of
           $\sigma(p\bar{p}\to WZ,ZZ) =  5.8^{+3.6}_{-3.0}~\text{pb}$
           (red dashed vertical line) corresponds to 
           a 95\% CL limit at $13~\text{pb}~(2.6 \times \sigma_\text{SM})$.}
 \label{fig:FC}
 \end{figure*}
To translate the result of our fit to the data to bounds or limits on
the  cross section of $WZ/ZZ$ production, we construct 
Feldman-Cousins bands by analyzing the distribution of fitted
(\textit{i.e.}, measured) cross sections in pseudo-experiments
generated with a variety of scale factors on the input signal cross
section~\cite{FC}. When running pseudo-experiments, we consider the
effect of additional systematic uncertainties that affect our
acceptance. These are, in order of increasing significance: jet
energy resolution ($0.7\%$), $\mett$ modeling ($1.0\%$), parton 
distribution functions ($2.0\%$), initial and final state radiation ($2.4\%$), and
luminosity and trigger efficiency uncertainties ($6.4\%$).
The set of input cross sections in our pseudo-experiments range from
$0.1$ to $3.0$ times the standard model value with a step size of
$0.1$. Fig.~\ref{fig:FC} shows the results of our Feldman-Cousins
analysis.
Based on a Monte Carlo simulation, the acceptance times efficiency for
the $WZ$ and $ZZ$ production is 4.1$\%$, and 4.6$\%$, respectively.


Our measured result, using the $1\sigma$ bands from the Feldman-Cousins 
analysis, is $\sigma(p\bar{p}\to WZ,ZZ) =
5.8^{+3.6}_{-3.0}~\text{pb}$, in agreement with the standard model 
prediction $\sigma_\text{SM}=5.1$~pb (\cite{campbell}). We perform
pseudo-experiments to calculate the probability (p-value) that the 
background-only model fluctuates up to the observed result (observed
p-value) and up to the median expected $s+b$ result (expected p-value). We 
observe a p-value of $2.7\%$, corresponding to a signal significance of 
$1.9\sigma$ where $1.7\sigma$ is expected. We set a limit on
$\sigma_{\scriptscriptstyle WZ,ZZ}<13~\mathrm{pb}~
(2.6\times \sigma_\text{SM})$ with 95\% C.L.
The techniques used here, in particular the $b$ tagging algorithm, are 
being integrated in the current
generation of searches for a low-mass Higgs boson.

\begin{acknowledgments}
We thank the Fermilab staff and the technical staffs of the
participating institutions for their vital contributions. This work
was supported by the U.S. Department of Energy and National Science
Foundation; the Italian Istituto Nazionale di Fisica Nucleare; the
Ministry of Education, Culture, Sports, Science and Technology of
Japan; the Natural Sciences and Engineering Research Council of
Canada; the National Science Council of the Republic of China; the
Swiss National Science Foundation; the A.P. Sloan Foundation; the
Bundesministerium f\"ur Bildung und Forschung, Germany; the Korean
World Class University Program, the National Research Foundation of
Korea; the Science and Technology Facilities Council and the Royal
Society, UK; the Institut National de Physique Nucleaire et Physique
des Particules/CNRS; the Russian Foundation for Basic Research; the
Ministerio de Ciencia e Innovaci\'{o}n, and Programa
Consolider-Ingenio 2010, Spain; the Slovak R\&D Agency; the Academy of
Finland; and the Australian Research Council (ARC). 
\end{acknowledgments}
  
\bibliography{prl}

\begin{thebibliography}{29}
\expandafter\ifx\csname natexlab\endcsname\relax\def\natexlab#1{#1}\fi
\expandafter\ifx\csname bibnamefont\endcsname\relax
  \def\bibnamefont#1{#1}\fi
\expandafter\ifx\csname bibfnamefont\endcsname\relax
  \def\bibfnamefont#1{#1}\fi
\expandafter\ifx\csname citenamefont\endcsname\relax
  \def\citenamefont#1{#1}\fi
\expandafter\ifx\csname url\endcsname\relax
  \def\url#1{\texttt{#1}}\fi
\expandafter\ifx\csname urlprefix\endcsname\relax\def\urlprefix{URL }\fi
\providecommand{\bibinfo}[2]{#2}
\providecommand{\eprint}[2][]{\url{#2}}

\bibitem[{\citenamefont{Hagiwara et~al.}(1987)}]{hagiwara}
\bibinfo{author}{\bibfnamefont{K.}~\bibnamefont{Hagiwara}}
  \bibnamefont{et~al.}, \bibinfo{journal}{Nucl. Phys. \textbf{B}}
  \textbf{\bibinfo{volume}{282}}, \bibinfo{pages}{253} (\bibinfo{year}{1987}).

\bibitem[{\citenamefont{Kober et~al.}(2007)\citenamefont{Kober, Koch, and
  Bleicher}}]{kober}
\bibinfo{author}{\bibfnamefont{M.}~\bibnamefont{Kober}},
  \bibinfo{author}{\bibfnamefont{B.}~\bibnamefont{Koch}}, \bibnamefont{and}
  \bibinfo{author}{\bibfnamefont{M.}~\bibnamefont{Bleicher}},
  \bibinfo{journal}{Phys. Rev. D} \textbf{\bibinfo{volume}{76}},
  \bibinfo{pages}{125001} (\bibinfo{year}{2007}).

\bibitem[{\citenamefont{Eichten et~al.}(2011)\citenamefont{Eichten, Lane, and
  Martin}}]{technicolor}
\bibinfo{author}{\bibfnamefont{E.~J.} \bibnamefont{Eichten}},
  \bibinfo{author}{\bibfnamefont{K.}~\bibnamefont{Lane}}, \bibnamefont{and}
  \bibinfo{author}{\bibfnamefont{A.}~\bibnamefont{Martin}},
  \bibinfo{journal}{Phys. Rev. Lett.} \textbf{\bibinfo{volume}{106}},
  \bibinfo{pages}{251803} (\bibinfo{year}{2011}).

\bibitem[{\citenamefont{Aaltonen et~al.}(2010{\natexlab{a}})}]{higgs_metbb}
\bibinfo{author}{\bibfnamefont{T.}~\bibnamefont{Aaltonen}} \bibnamefont{et~al.}
  (\bibinfo{collaboration}{CDF Collaboration}), \bibinfo{journal}{Phys. Rev.
  Lett.} \textbf{\bibinfo{volume}{104}}, \bibinfo{pages}{141801}
  (\bibinfo{year}{2010}{\natexlab{a}}).

\bibitem[{\citenamefont{Aaltonen et~al.}(2010{\natexlab{b}})}]{cdf_ww_dil}
\bibinfo{author}{\bibfnamefont{T.}~\bibnamefont{Aaltonen}} \bibnamefont{et~al.}
  (\bibinfo{collaboration}{CDF Collaboration}), \bibinfo{journal}{Phys. Rev.
  Lett.} \textbf{\bibinfo{volume}{104}}, \bibinfo{pages}{201801}
  (\bibinfo{year}{2010}{\natexlab{b}}).

\bibitem[{\citenamefont{Abazov et~al.}(2009)}]{d0_ww_dil}
\bibinfo{author}{\bibfnamefont{V.~M.} \bibnamefont{Abazov}}
  \bibnamefont{et~al.} (\bibinfo{collaboration}{D0 Collaboration}),
  \bibinfo{journal}{Phys. Rev. Lett.} \textbf{\bibinfo{volume}{103}},
  \bibinfo{pages}{191801} (\bibinfo{year}{2009}).

\bibitem[{\citenamefont{Aaltonen et~al.}(2009{\natexlab{a}})}]{metjj_prl}
\bibinfo{author}{\bibfnamefont{T.}~\bibnamefont{Aaltonen}} \bibnamefont{et~al.}
  (\bibinfo{collaboration}{CDF Collaboration}), \bibinfo{journal}{Phys. Rev.
  Lett.} \textbf{\bibinfo{volume}{103}}, \bibinfo{pages}{091803}
  (\bibinfo{year}{2009}{\natexlab{a}}).

\bibitem[{\citenamefont{Aaltonen et~al.}(2010{\natexlab{c}})}]{lmetjj_prl}
\bibinfo{author}{\bibfnamefont{T.}~\bibnamefont{Aaltonen}} \bibnamefont{et~al.}
  (\bibinfo{collaboration}{CDF Collaboration}), \bibinfo{journal}{Phys. Rev.
  Lett.} \textbf{\bibinfo{volume}{104}}, \bibinfo{pages}{101801}
  (\bibinfo{year}{2010}{\natexlab{c}}).

\bibitem[{wwc()}]{wwcomment}
\bibinfo{note}{This last sample also contains $WW$ events}.

\bibitem[{met()}]{metdef}
\bibinfo{note}{We define the missing transverse momentum $\vec{\mett}$$\equiv
  -\sum_i E_{\rm T}^i {\bf n}_i$, where ${\bf n}_i$ is the unit vector in the
  azimuthal plane that points from the beamline to the $i$th calorimeter tower.
  We call the magnitude of this vector \mett}.

\bibitem[{\citenamefont{Abulencia et~al.}(2007)}]{CDF_detect_A}
\bibinfo{author}{\bibfnamefont{A.}~\bibnamefont{Abulencia}}
  \bibnamefont{et~al.} (\bibinfo{collaboration}{CDF Collaboration}),
  \bibinfo{journal}{J. Phys. G} \textbf{\bibinfo{volume}{34}},
  \bibinfo{pages}{2457} (\bibinfo{year}{2007}).

\bibitem[{\citenamefont{Abe et~al.}(1992)}]{jetclu}
\bibinfo{author}{\bibfnamefont{F.}~\bibnamefont{Abe}} \bibnamefont{et~al.}
  (\bibinfo{collaboration}{CDF Collaboration}), \bibinfo{journal}{Phys. Rev. D}
  \textbf{\bibinfo{volume}{45}}, \bibinfo{pages}{1448} (\bibinfo{year}{1992}).

\bibitem[{\citenamefont{Bhatti et~al.}(2006)}]{jesnim}
\bibinfo{author}{\bibfnamefont{A.}~\bibnamefont{Bhatti}} \bibnamefont{et~al.},
  \bibinfo{journal}{Nucl. Instrum. Methods A} \textbf{\bibinfo{volume}{566}},
  \bibinfo{pages}{375} (\bibinfo{year}{2006}).

\bibitem[{whi()}]{whichjets}
\bibinfo{note}{All jets in the region $|\eta|<3.6$ and $E_T>5~\mathrm{GeV}$ are
  considered.}

\bibitem[{\citenamefont{Goncharov et~al.}(2006)}]{emtiming}
\bibinfo{author}{\bibfnamefont{M.}~\bibnamefont{Goncharov}}
  \bibnamefont{et~al.}, \bibinfo{journal}{Nucl. Instrum. Methods A}
  \textbf{\bibinfo{volume}{565}}, \bibinfo{pages}{543} (\bibinfo{year}{2006}).

\bibitem[{\citenamefont{Freeman et~al.}(2011)\citenamefont{Freeman, Ketchum,
  Lewis, Poprocki, Pronko, Rusu, and Wittich}}]{nim_bness}
\bibinfo{author}{\bibfnamefont{J.}~\bibnamefont{Freeman}},
  \bibinfo{author}{\bibfnamefont{W.}~\bibnamefont{Ketchum}},
  \bibinfo{author}{\bibfnamefont{J.}~\bibnamefont{Lewis}},
  \bibinfo{author}{\bibfnamefont{S.}~\bibnamefont{Poprocki}},
  \bibinfo{author}{\bibfnamefont{A.}~\bibnamefont{Pronko}},
  \bibinfo{author}{\bibfnamefont{V.}~\bibnamefont{Rusu}}, \bibnamefont{and}
  \bibinfo{author}{\bibfnamefont{P.}~\bibnamefont{Wittich}},
  \bibinfo{journal}{Nucl. Instrum. Methods A} \textbf{\bibinfo{volume}{663}},
  \bibinfo{pages}{37} (\bibinfo{year}{2011}), \eprint{arXiv:1108.4738v2}.

\bibitem[{sig()}]{signif}
\bibinfo{note}{The significance of a quantity is defined as the value divided
  by its uncertainty.}

\bibitem[{\citenamefont{Aaltonen et~al.}(2009{\natexlab{b}})}]{cdf_smt}
\bibinfo{author}{\bibfnamefont{T.}~\bibnamefont{Aaltonen}} \bibnamefont{et~al.}
  (\bibinfo{collaboration}{CDF Collaboration}), \bibinfo{journal}{Phys. Rev. D}
  \textbf{\bibinfo{volume}{79}}, \bibinfo{pages}{052007}
  (\bibinfo{year}{2009}{\natexlab{b}}).

\bibitem[{\citenamefont{Sj\"ostrand et~al.}(2006)}]{pythia}
\bibinfo{author}{\bibfnamefont{T.}~\bibnamefont{Sj\"ostrand}}
  \bibnamefont{et~al.}, \bibinfo{journal}{J. High Energy Phys.}
  \textbf{\bibinfo{volume}{05}}, \bibinfo{pages}{026} (\bibinfo{year}{2006}).

\bibitem[{\citenamefont{Mangano et~al.}(2003)}]{alpgen}
\bibinfo{author}{\bibfnamefont{M.~L.} \bibnamefont{Mangano}}
  \bibnamefont{et~al.}, \bibinfo{journal}{J. High Energy Phys.}
  \textbf{\bibinfo{volume}{07}}, \bibinfo{pages}{001} (\bibinfo{year}{2003}).

\bibitem[{\citenamefont{Alwall et~al.}(2007)}]{madgraph}
\bibinfo{author}{\bibfnamefont{J.}~\bibnamefont{Alwall}} \bibnamefont{et~al.},
  \bibinfo{journal}{J. High Energy Phys.} \textbf{\bibinfo{volume}{09}},
  \bibinfo{pages}{028} (\bibinfo{year}{2007}).

\bibitem[{\citenamefont{Brun et~al.}(1978)}]{geant3}
\bibinfo{author}{\bibfnamefont{R.}~\bibnamefont{Brun}} \bibnamefont{et~al.}
  (\bibinfo{year}{1978}), \bibinfo{note}{{CERN} Report CERN-DD-78-2-REV
  (unpublished)}.

\bibitem[{\citenamefont{Campbell and Ellis}(1999)}]{campbell}
\bibinfo{author}{\bibfnamefont{J.~M.} \bibnamefont{Campbell}} \bibnamefont{and}
  \bibinfo{author}{\bibfnamefont{R.~K.} \bibnamefont{Ellis}},
  \bibinfo{journal}{Phys. Rev. D} \textbf{\bibinfo{volume}{60}},
  \bibinfo{pages}{113006} (\bibinfo{year}{1999}).

\bibitem[{\citenamefont{Pumplin et~al.}(2002)}]{cteq6m}
\bibinfo{author}{\bibfnamefont{J.}~\bibnamefont{Pumplin}} \bibnamefont{et~al.},
  \bibinfo{journal}{J. High Energy Phys.} \textbf{\bibinfo{volume}{0207}},
  \bibinfo{pages}{012} (\bibinfo{year}{2002}).

\bibitem[{\citenamefont{Aaltonen et~al.}(2010{\natexlab{d}})}]{mclimit}
\bibinfo{author}{\bibfnamefont{T.}~\bibnamefont{Aaltonen}} \bibnamefont{et~al.}
  (\bibinfo{collaboration}{CDF Collaboration}), \bibinfo{journal}{Phys. Rev. D}
  \textbf{\bibinfo{volume}{82}}, \bibinfo{pages}{112005}
  (\bibinfo{year}{2010}{\natexlab{d}}).

\bibitem[{\citenamefont{Langenfeld et~al.}(2009)\citenamefont{Langenfeld, Moch,
  and Uwer}}]{ttbar-xsec}
\bibinfo{author}{\bibfnamefont{U.}~\bibnamefont{Langenfeld}},
  \bibinfo{author}{\bibfnamefont{S.}~\bibnamefont{Moch}}, \bibnamefont{and}
  \bibinfo{author}{\bibfnamefont{P.}~\bibnamefont{Uwer}},
  \bibinfo{journal}{Phys. Rev. D} \textbf{\bibinfo{volume}{80}},
  \bibinfo{pages}{054009} (\bibinfo{year}{2009}).

\bibitem[{\citenamefont{Kidonakis}(2006)}]{singlet-t}
\bibinfo{author}{\bibfnamefont{N.}~\bibnamefont{Kidonakis}},
  \bibinfo{journal}{Phys. Rev. D} \textbf{\bibinfo{volume}{74}},
  \bibinfo{pages}{114012} (\bibinfo{year}{2006}).

\bibitem[{\citenamefont{Kidonakis}(2010)}]{singlet-s}
\bibinfo{author}{\bibfnamefont{N.}~\bibnamefont{Kidonakis}},
  \bibinfo{journal}{PoS(DIS2010)} \textbf{\bibinfo{volume}{196}}
  (\bibinfo{year}{2010}), \eprint{arXiv:1005.3330v1}.

\bibitem[{\citenamefont{Feldman and Cousins}(1998)}]{FC}
\bibinfo{author}{\bibfnamefont{G.~J.} \bibnamefont{Feldman}} \bibnamefont{and}
  \bibinfo{author}{\bibfnamefont{R.~D.} \bibnamefont{Cousins}},
  \bibinfo{journal}{Phys. Rev. D} \textbf{\bibinfo{volume}{57}},
  \bibinfo{pages}{3873} (\bibinfo{year}{1998}).

\end{thebibliography}

\end{document}